\documentclass[11pt]{article}
\usepackage{moriond,epsfig}
\pdfoutput=1

\def\be{\begin{equation}}
\def\ee{\end{equation}}
\def\bea{\begin{eqnarray}}
\def\eea{\end{eqnarray}}

\begin{document}
\vspace*{0cm}
\title{$W^+W^+jj$ at NLO in QCD: an exotic Standard Model signature at the LHC}

\author{ Tom Melia }

\address{Department of Physics, Theoretical Physics, 1 Keble Road,\\
Oxford OX1 3NP, England}

\maketitle\abstracts{
The process $pp \to W^+W^+jj$ gives rise to an exotic Standard Model signature
at the LHC, involving high-$p_{\perp}$ like-sign leptons, missing energy and jets. 
In this brief article the motivation for study, along with selected results from the computation of NLO QCD 
corrections to the QCD-mediated part of this process\cite{us} are presented. It is shown
that the corrections reduce the dependence of the cross-section on renormalisation 
and factorisation scales, and produce a relatively hard third jet in a significant
fraction of events.}

\section{Introduction}
\begin{figure}[t]
\begin{center}
\includegraphics[angle=0,scale=1.22]{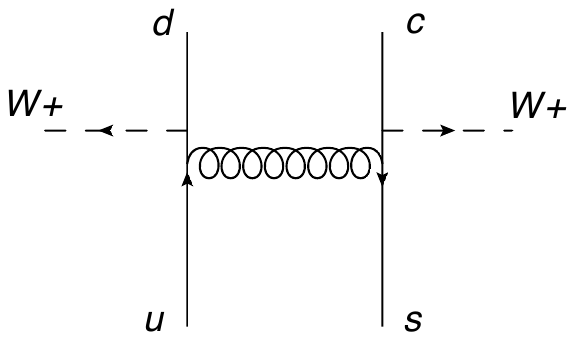}

\caption{A typical Feynman diagram which contributes to the process $pp \to W^+ W^+
jj$.}
\label{feyn}
\end{center}
\end{figure}

The process $pp\to W^+W^+jj$ is a quirky one, both theoretically and 
experimentally \cite{us}. At a particle collider, the signature involves 
like-sign leptons, jets and missing energy -- an exotic signal from the Standard
Model! 
The mechanisms by which two positively charged $W$ bosons can be created are rather restricted and this leads to the theoretical quirkiness.
Figure \ref{feyn} shows a typical Feynman diagram for this process (even though
 in our calculation we do not compute a single one).
Charge conservation requires that the $W$ bosons be emitted from 
separate quark lines. Because a massive particle is always produced on
each fermion line, the cross-section for the process $pp \to W^+ W^+ jj$ 
remains finite {\it even if the requirement that two jets are observed is
lifted}. This rather unusual feature is seldom present in NLO
QCD calculations.

At $\sqrt{s} = 14~{\rm TeV}$, the cross-section for this process
is about $1~{\rm pb}$ (40\% of this for $W^-W^-jj$) and therefore accessible. When the $W$ 
bosons decay leptonically, the two positively charged isolated leptons and
missing energy give rise to a nearly background-free signature.
The observation of this process is interesting in its own right, but
there are other reasons to study it. Of particular importance are
various physics cases for which $pp \to W^+ W^+ jj$ is a background
process.  Interestingly, such cases can be found both within and
beyond the Standard Model.  For example, it is possible to use
same-sign lepton pairs to study double parton scattering at the
LHC \cite{dps} in which case the single scattering process $pp \to W^+
W^+ jj$ is the background. Events with same-sign leptons, missing
energy and two jets can also appear due to resonant slepton production
which may occur in $R$-parity violating SUSY models \cite{dreiner} or
in the case of diquark production \cite{berger} with subsequent decay of
the diquark to e.g. pairs of top quarks. Similarly, one of the
possible production mechanisms of the double-charged Higgs boson at
the LHC has a signature of two same-sign leptons, missing energy and
two jets \cite{maalampi}. A final reason is that the diagrams which contribute
 are a subset of the diagrams for $pp\to W^+W^-jj$, an important
background to Higgs boson production in weak boson fusion. This calculation can be seen as a theoretical stepping stone
leading to the recently computed NLO corrections to $pp\to W^+W^-jj$ \cite{us2}.

 Full details of 
the calculation can be found in Ref. \cite{us}, but it is worth pointing out
that because $pp \to W^+ W^+ jj$ is a $2 \to
4$ process, one-loop six-point tensor
integrals of relatively high rank need to be dealt with. 
It is only very recently that theoretical methods for one-loop calculations
have become adequate to handle computations of such a complexity.
We use the framework of generalized $D$-dimensional unitarity
{\cite{egk,Giele:2008ve}}, closely following and extending the
implementation described in Ref.~{\cite{Ellis:2008qc}},and  demonstrating the
ability of this method to deal with complicated final states involving two colourless
particles. This process has since been implemented in the {\tt POWHEG BOX} \cite{us3}, and is the first
$2\to4$ NLO process to be matched with a parton shower.

\section{Results}
We consider proton-proton collisions at a center-of-mass energy
$\sqrt{s} = 14~ {\rm TeV}$. We require leptonic decays of the
$W$-bosons and consider the final state $e^+ \mu^+ \nu_e \nu_\mu $ . The
$W$-bosons are on the mass-shell and we neglect quark flavour mixing.
We impose standard cuts on lepton transverse momenta $p_{\perp, l} >
20~{\rm GeV}$, missing transverse momentum $p_{\perp, \rm miss} >
30~{\rm GeV}$ and charged lepton rapidity $| \eta_l| < 2.4$. We define
jets using anti-$k_{\perp}$ algorithm \cite{Cacciari:2008gp}, with
$\Delta R_{j_1j_2} = 0.4$ and, unless otherwise specified, with a
transverse momentum cut $p_{\perp, j} = 30~{\rm GeV}$ on the two
jets. The mass of the $W$-boson is taken to be $m_W = 80.419~{\rm
  GeV}$, the width $\Gamma_W = 2.140$~{\rm GeV}. $W$ couplings to
fermions are obtained from $\alpha_{\rm QED} (m_Z) = 1 / 128.802$ and
$\sin^2 \theta_W = 0.2222$.  We use MSTW08LO parton distribution
functions for leading order and MSTW08NLO for next-to-leading order
computations, corresponding to $\alpha_s(M_Z) = 0.13939$ and
$\alpha_s(M_Z) = 0.12018$ respectively~\cite{Martin:2009iq}. We do not
impose lepton isolation cuts. All results discussed below apply to the
QCD production $pp \to W^+ W^+ jj$; the electroweak contribution to
this process is ignored.

\begin{figure}[t]
\begin{center}
\includegraphics[angle=0,scale=0.55]{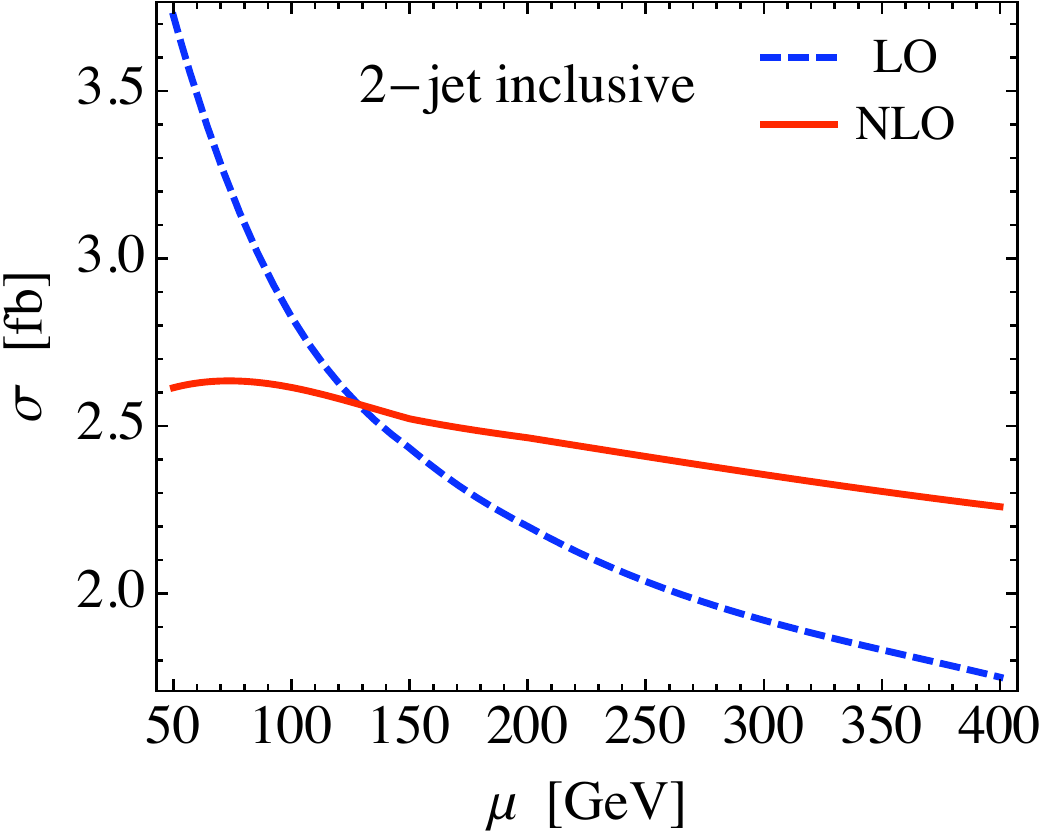}
\includegraphics[angle=0,scale=0.53]{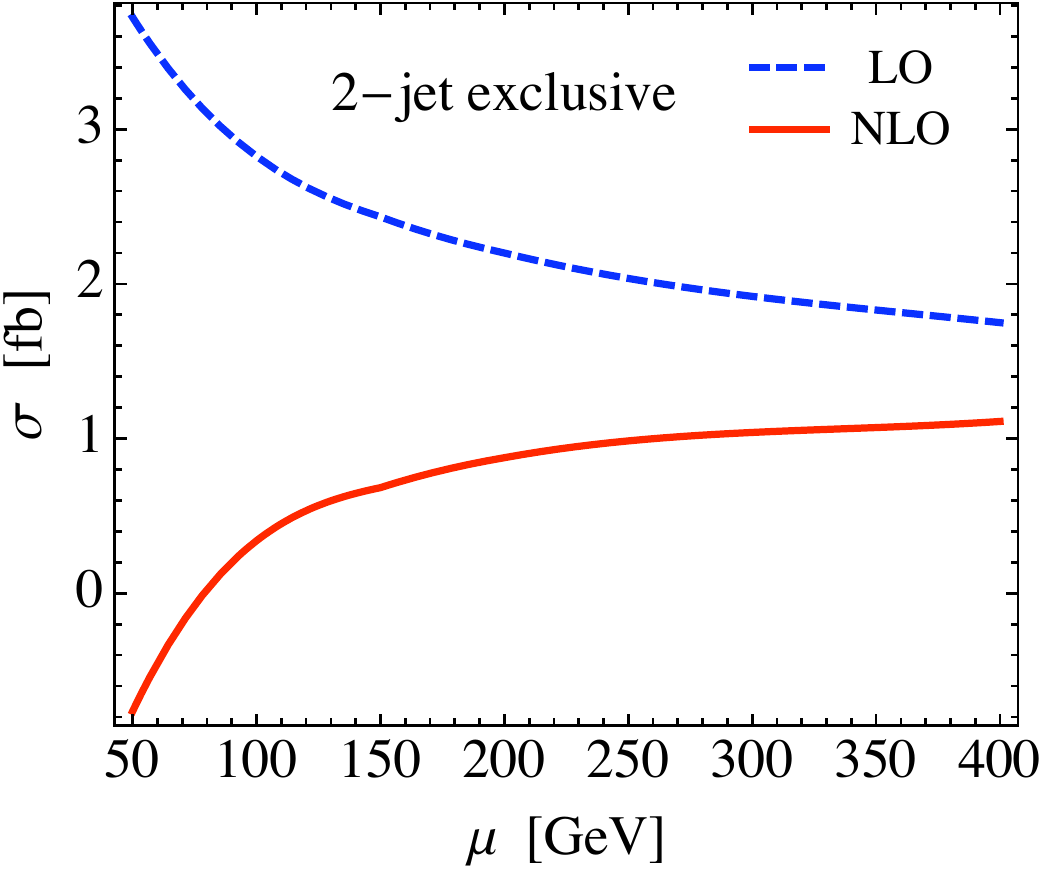}

\vspace{0.1cm}

\includegraphics[angle=0,scale=0.55]{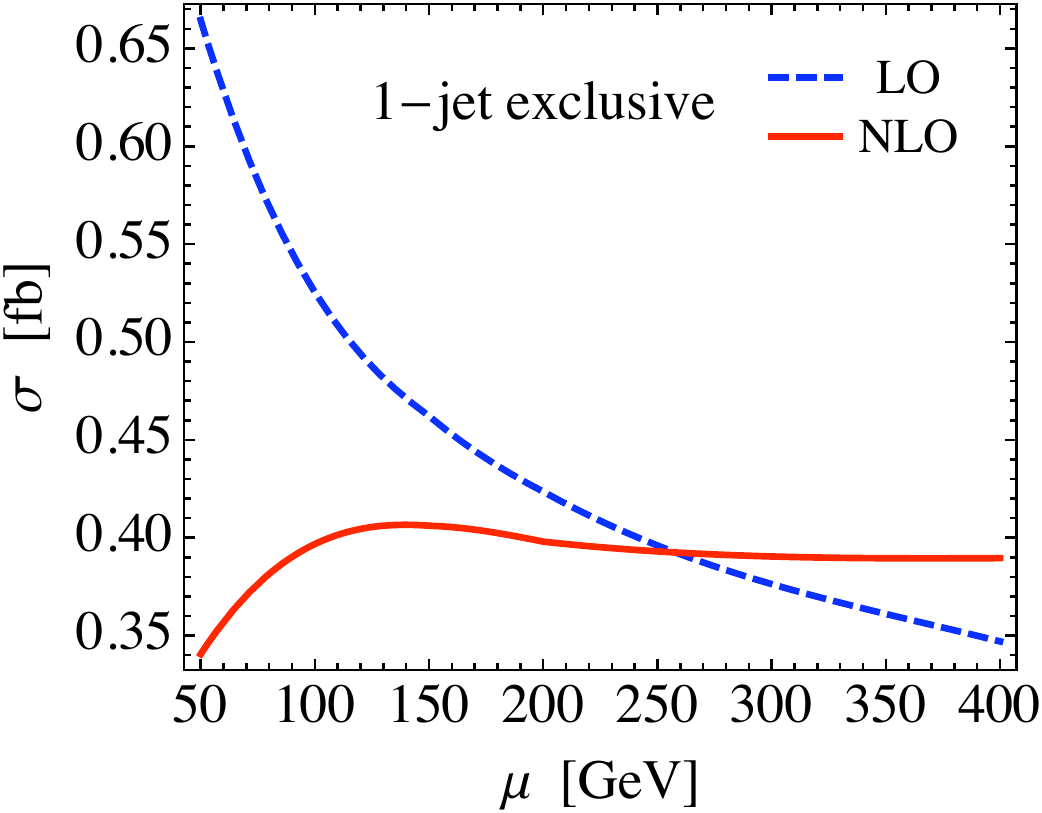}
\includegraphics[angle=0,scale=0.55]{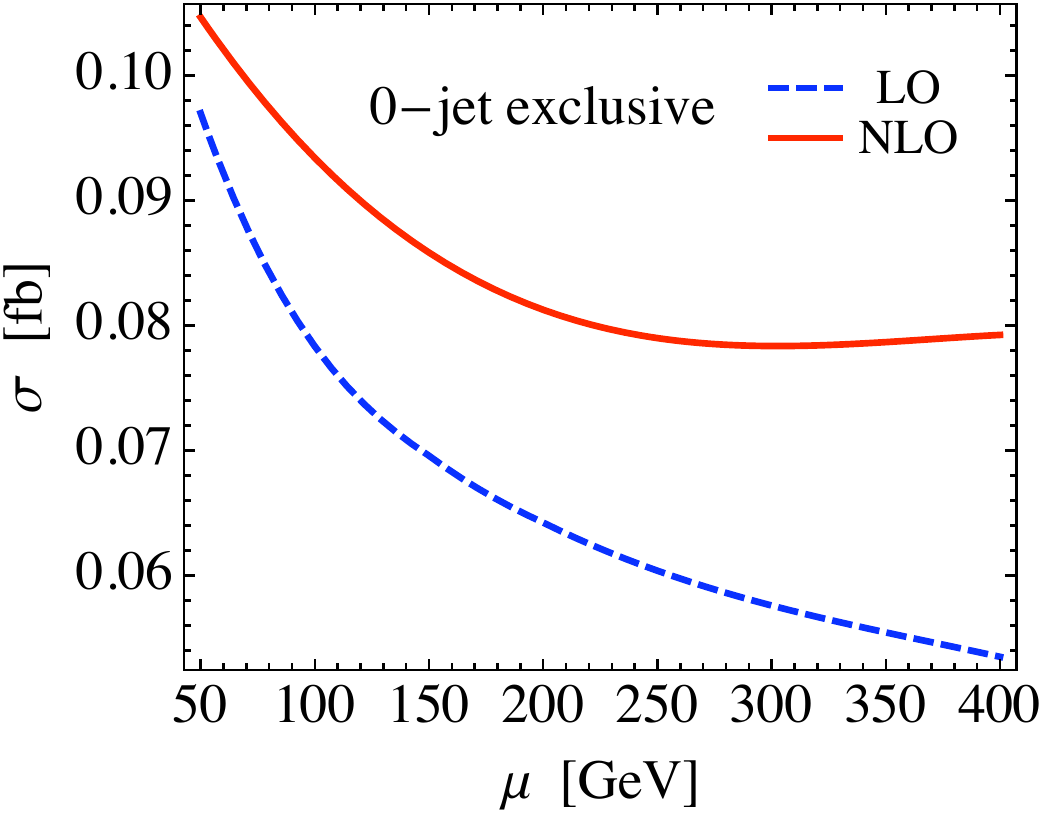}
\caption{The dependence on factorisation and renormalisation scales of
  cross-sections for $pp \to e^+\, \mu^+\, {\nu}_{e}\, {\nu}_{\mu} + n~{\rm
    jets}$, $n = 0,1,2$ at leading and next-to-leading order in
  perturbative QCD. Here $\mu_{\rm F} = \mu_{\rm R} = \mu$. }
\label{fig1}
\end{center}
\end{figure}
Since the cross section remains finite
even if the requirement that two jets are observed is lifted, we can consider
 the production of same-sign gauge
bosons in association with $n$ jets $pp \to W^+ W^+ + n~{\rm jets}$,
where $n = 0, 1, 2$ or $n \ge 2$. 
Fig.~\ref{fig1} shows the dependence of the
production cross-sections for $pp \to e^+ \mu^+ \nu_{e} \nu_{\mu} +
n~{\rm jets}$ on the renormalisation and factorisation scales, which
we set equal to each other.
\begin{figure}[t]
\begin{center}
\includegraphics[angle=0,scale=0.47]{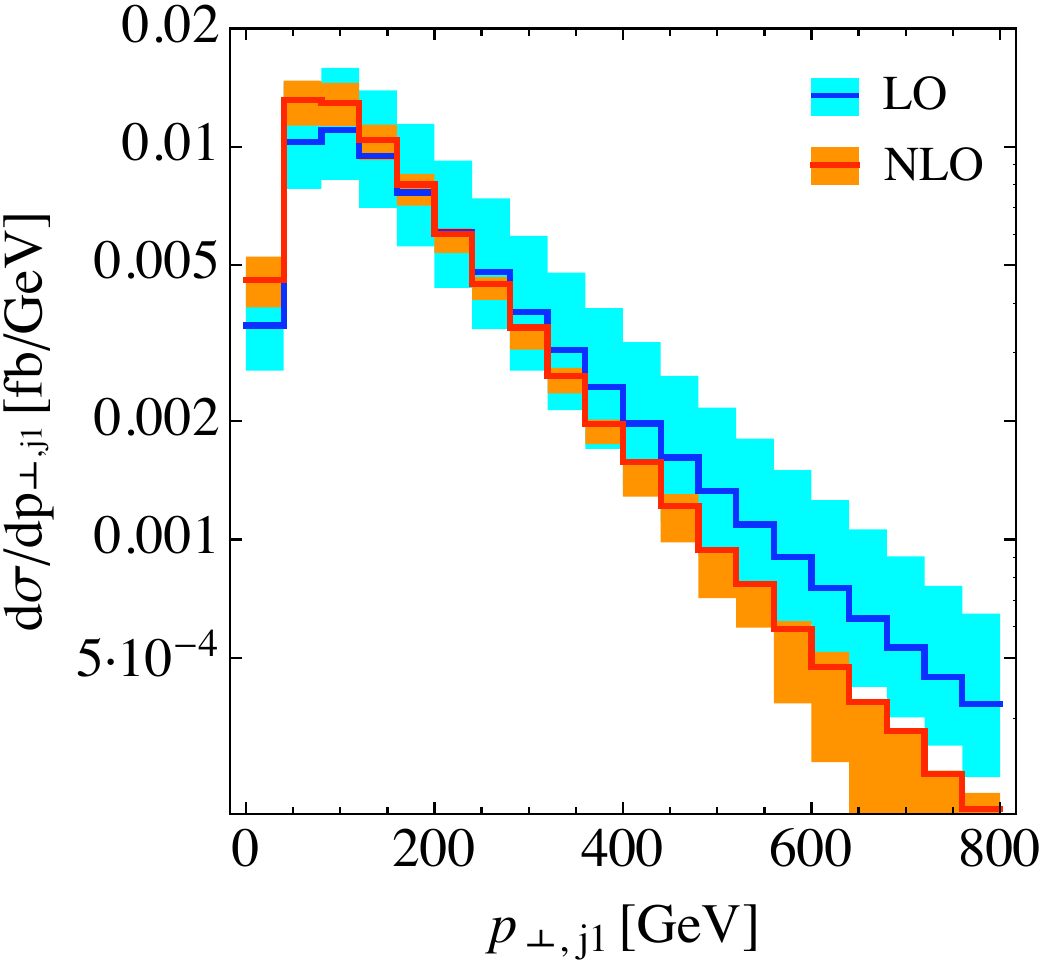}
\qquad
\includegraphics[angle=0,scale=0.45]{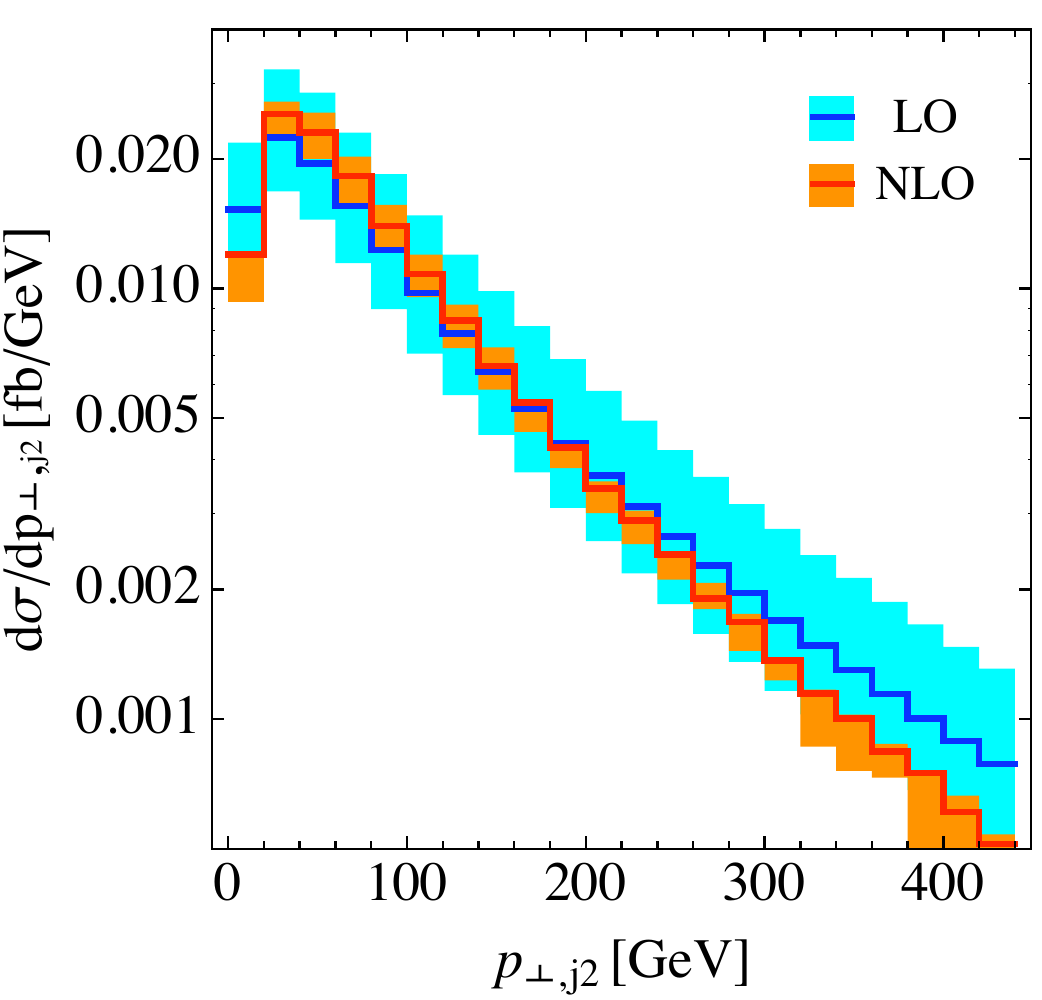}

\vspace{0cm}

\includegraphics[angle=0,scale=0.45]{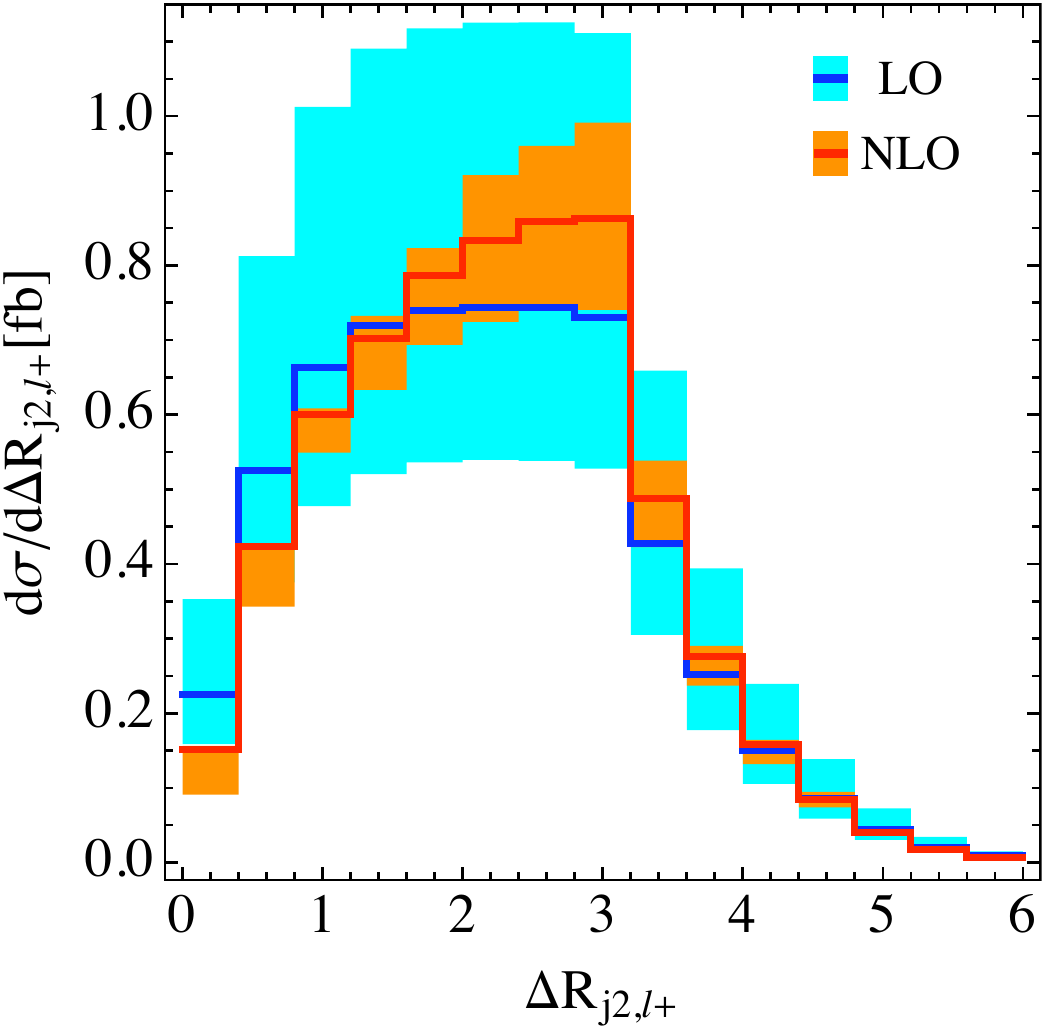}
\qquad 
\includegraphics[angle=0,scale=0.45]{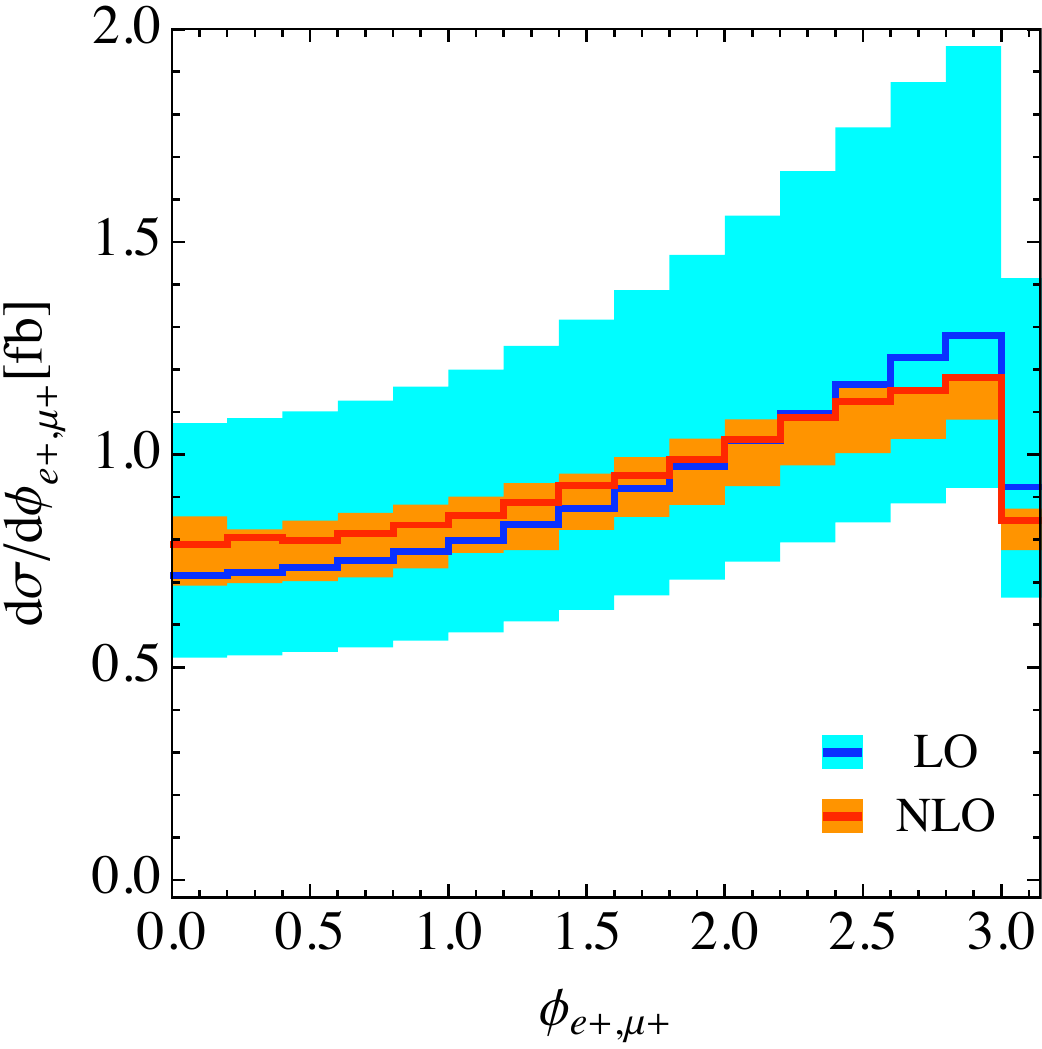}
\caption{Distributions of the transverse momentum of the 
two hardest jets and angular distributions in the
  process $pp \to e^+\, \mu^+\, {\nu}_{e}\, {\nu}_{\mu}\, + 2~{\rm jets}$ at
  leading and next-to-leading order in perturbative QCD for inclusive
  two-jet events.  The bands show renormalisation and factorisation
  scale uncertainty, for $50~{\rm GeV} \leq \mu \leq 400~{\rm
    GeV}$. Solid lines show leading and next-to-leading order
  predictions for $\mu = 150~{\rm GeV}$. }
\label{fig2}
\end{center}
\end{figure}
Considering the
range of scales $50~{\rm GeV} \le \mu \le 400~{\rm GeV}$, we find the
two-jet inclusive cross-section to be $\sigma^{\rm LO} = 2.7 \pm
1.0~{\rm fb}$ at leading order and $\sigma^{\rm NLO} = 2.44 \pm
0.18~{\rm fb}$ at next-to-leading order. The forty percent scale
uncertainty at leading order is reduced to less than ten percent at
NLO.  We observe similar stabilization of the scale dependence for the
$0$- and $1$-jet exclusive multiplicities.  Combining these
cross-sections we obtain a total NLO cross-section of about $2.90~{\rm
  fb}$ for $pp \to e^+ \mu^+ \nu_e \nu_\mu $ inclusive
production. This implies about $60$ $e^+\mu^++ e^+e^+ + \mu^+\mu^+$
events per year at the LHC with $10~{\rm fb}^{-1}$ annual
luminosity. While this is not a gigantic number, such events will have
a very distinct signature, so they will definitely be seen and it will
be possible to study them. 

The dramatic change in the two-jet exclusive cross-section apparent from 
 Fig.~\ref{fig1} is discussed and investigated in Ref.\cite{us}. We find that 
 the feature observed here, that the two-jet exclusive is significantly smaller
 than the two-jet inclusive, remains present when we increase the jet cut and 
 so allow for greater perturbative convergence of the exclusive cross section. 
 This smallness implies that quite
a large fraction of events in $pp \to e^+\mu^+ \nu_e \nu_\mu + \ge
2~{\rm jets}$ have a relatively hard third jet. This feature may be
useful for rejecting contributions of $pp \to W^+ W^+ jj$ when looking
for multiple parton scattering.

%This NLO calculation has been implemented in the POWHEG BOX, and 
%is the first $2\to4$ process calculated in a NLO + parton shower framework.
%Figure \ref{fig2} shows selected kinematic distributions for a pure NLO and
%for events showered with PYTHIA.

Selected kinematic distributions are shown in Fig.~\ref{fig2}. It is clear 
that jets in $pp \to
W^+ W^+ jj$ are hard; a typical transverse momentum of the hardest jet
is close to $100~{\rm GeV}$ and the transverse momentum of the
next-to-hardest jet is close to $40~{\rm GeV}$.  The NLO distributions show 
a characteristic depletion at large
values of $p_{\perp,j}$.  One reason this change occurs is because a
constant, rather than a dynamical, renormalisation scale is used in
our leading order calculation. Scale dependencies of
the distributions are reduced dramatically.
The angular distance 
$\Delta R_{lj}$ between a charged
lepton of fixed flavor ($e^+$ or $\mu^+$) and the next-to-hardest jet is displayed, 
as well as the distribution of the
relative azimuthal angle of the two charged leptons. Although the
distribution of angular distance between leptons and the next-to-hardest jet is broad,
it peaks at $\Delta R_{lj} \approx 3$.  NLO QCD effects do not change this conclusion
but, interestingly, they make the angular distance between
next-to-hardest jet and the charged lepton somewhat larger. 
The distribution of the relative azimuthal
angle of the two charged leptons becomes less peaked at $\Delta
\phi_{l^+ l^+} = \pi$, although the two leptons still prefer to be
back to back.
It is interesting to remark that, if the two same sign leptons are
produced through a double-parton scattering mechanism, their
directions are not correlated. Hence, yet another possibility to
reduce the single-scattering-background is to cut on the relative
azimuthal angle between the two leptons.

To conclude, selected results from the calculation of NLO QCD corrections to the QCD-mediated process
$pp\to W^+W^+jj$ have been presented. Methods developed very recently 
for computing one-loop amplitudes allowed for relatively straightforward calculation of this $2 \to 4$ process.
The detector signature
is an exotic one, and is exciting to study at the LHC.

\section*{Acknowledgments}
I wish to thank the organisers of the Moriond QCD session for providing
both financial support and a fantastic conference. This write-up is based on
work done in collaboratoration with Kirill Melnikov, Raoul R\"ontsch and Giulia
Zanderighi and closely follows Ref.\cite{us}, and the research is supported by the British Science and Technology Facilities Council.

\end{document}